\documentstyle[11pt]{article}
\title{
\hfill\parbox{4cm}{\normalsize LPHEA /00-05}\\
\hfill\parbox{4cm}{\normalsize UFR-HEP/00-22}\\
\vspace{0.5cm}
{\small {\bf Heavy-meson masses via Dick interquark potential}}}
\author{T. Barakat \thanks{e-mail: zayd95@hotmail.com} \\
{\it {\small Near East University, Lefko\c{s}a, P.O. BOX 1033
Mersin 10 - Turkey }}\and \\
M. Chabab \thanks{e-mail:mchabab@ucam.ac.ma}  \\
{\it {\small LPHEA, Physics Department, Faculty of Science-Semlalia,
Cadi Ayyad University}}\\
{\it{\small  P.O. Box 2390, Marrakesh 40001, Morocco}}\\
{\it {\small and}}\\
{\it {\small UFR-PHE, Physics Department, Faculty of Science, Rabat, Morocco }}}
\date{}
\begin{document}
\begin{titlepage}
\maketitle
\begin{abstract}
\baselineskip .5cm
We study Dick quark-antiquark potential (up to a color factor)
$V_D(r)={-\alpha_s \over r}+gf\sqrt{N_c\over{2(N_c-1)}}\ln[exp(2mr)-1]$ in the
 heavy meson sector. This potential emerges from an effective dilaton-gluon
coupling inspired from string theory and proves to be linearly rising at large
distances which ensures quark confinement. The semi-relativistic wave
 equation which appears in the theory of relativistic quark-antiquark bound
states is used. This equation is cast into a constituent second order
Schr\"{o}dinger-like equation with the inclusion of relativistic corrections
up to order $(v/c)^{2}$ in the quark speeds. The resulting equation is solved
for Dick potential via the Shifted-$l$ expansion technique (SLET). The
obtained results show that the spin-averaged energy levels of heavy-mesons are
well explained and agree with other potential models or QCD sum rules
predictions.  Moreover, as a by-product, our analysis assign to the dilaton a
mass around 56.9 MeV lying within the range of many theoretical scenario of
dilatonic mass.  \\
\end{abstract}
\end{titlepage}
\textwidth 13.3cm
\parskip .5cm
\baselineskip .85cm
\textheight 18.40cm
\section{ Introduction.}
\hspace{0.6cm}
The problem of calculating spin-averaged energy levels heavy-mesons is a very
old subject, but still an important theme in the existing literature. At
present muchexperimental material on the masses of ground and excited states of heavy quarkonia has been accumulated. By comparing theoretical predictions with experimental data, one can obtain information on the form of the potential of
the quark-antiquark interaction. This information illuminates the most
important features of quantum chromodynamics (QCD) dynamics: the asymptotic
freedom and confinement. At present it is not possible to obtain the
potential of the interquark interaction in the whole range of distances from
the basic principles of QCD. Consequently, the corresponding interquark
potential has to be determined phenomenologically. Therefore, to explain the
meson masses within the experimental limits, serious attempts to build a
quantitative potential model for quarks were made by many authors. The number
of papers on quark potential model is so large that one cannot quote even a
substantial fraction of them, but simply call attention to review [1], which
contains many references on the subject.

To understand confinement in gauge theories, much attention has been focused
on quark models based on QCD, because of their success in
understanding the spectroscopy of mesons, especially the new ones. At short
distances, QCD suggests a Coulomb-type potential $-\frac{\alpha_{s}}{r}$,
where $\alpha_{s}$ is the quark-gluon fine-structure constant. At long
distances one expects a confining potential. Lattice gauge theory [2], and
string models [3] lead one to expect a linear confining potential. Now it is
firmly established that the combination of linear confining potential, plus
coulombic-type short distance potential, plus one-gluon-exchange forces
provides a good fit to meson mass spectra [4]. However, one would be hard
pressed to say that they are well understood. Too many of the $Q\bar Q$
states predicted by the quark potential model have yet to be seen-only the
ground state S, and P wave multiplets are filled. Therefore, until
confinement is better understood, more states have been sited, and their
properties measured, we cannot say that the subject is completed.

 Recently, it has been pointed out by Dick [5,6,7] that an interquark
potential is constructed through the inclusion of a scalar field (dilaton) in
gauge theories. This potential proves to be rising at large distances
which ensures quark confinement, and at the same time it
incorporates the asymptotic freedom at small distances. Therefore, the
aim of this paper is to dedicate more efforts to understand this new
confinement generating mechanism through the investigation of the
phenomenological application of Dick interquark potential $V_{D}(r)$ in the
heavy mesons sector. This problem will be approached as in a previous work
[8]. Therein, it has been
demonstrated that the shifted-$l$ expansion (SLET), where $l$ is the angular
momentum, provides a powerful, systematic and analytic technique for
determining the bound states of the semi-relativistic wave equation
consisting of two quarks of masses $m_{1}$, $m_{2}$, and total energy $M$ in
any spherically symmetric potential, even one which has no small coupling
constant parameter. It simply consists of using $1/\bar{l}$ as a
pseudo-perturbation parameter, where $\bar{l}=l-\beta$, and $\beta$ is a
suitable shift. This shift is vital for it removes the poles that
would emerge, at lowest orbital states with $l=0$, in our proposed
expansion below. This method yields very accurate and rapidly converging
energy eigenvalue series. It also handles highly excited states which pose
roblems for variational methods [9]. Moreover, relativistic corrections are
included in a consistent way. In this spirit, this paper is organized as
follows: Section 2 recalls the conversion of the semi-relativistic equation
into an equivalent Schr\"{o}dinger- type equation, and there the Shifted-$l$
expansion technique (SLET) for this equation with any spherically symmetric
potential is introduced. Section 3 is devoted to describe the
phenomenological application of Dick quark-antiquark interaction potential.
We present, in this section, our numerical results of the spin-averaged
energy levels of charmonium, bottomonium and $b \bar c$ families, in
connection to the dilaton mass, and then we draw our conclusion.
\section{SLET for the semi-relativistic wave equation with any spherically
symmetric potential}
There are different methods to calculate the bound state energies of the
semi-relativistic wave equation [10], and references therein. However, the
method we present here is chosen for different reasons; ease of
implementation, high accuracy, and substantial computation time reduction is
achieved. An expansion in the powers of $(v/c)^{2}$ up to two terms in
the semi-relativistic equation which is a combination of relativistic
kinematics with some static interaction potential yields [8]:
\begin{equation}
\left[-\frac{1}{2\mu}\frac{d^{2}}{dr^{2}}+
\frac{l(l+1)}{2\mu r^{2}}+\gamma(r)+\frac{E_{n\ell}V(r)}{\eta}\right]R_{n\ell}
(r)=(\frac{E^{2}_{n\ell}}{2\eta}+E_{n\ell})R_{n\ell}(r),
\end{equation}
where $\gamma(r)=V(r)-V^{2}(r)/2\eta$, $E_{n\ell}=M-m_{1}-m_{2}$,
$\mu=m_{1}m_{2}/(m_{1}+m_{2})$ is the reduced mass, and $\eta=\nu/\mu^{2}$,
where $\nu$ is a useful parameter defined as
$\nu=m^{3}_{1}m^{3}_{2}/(m^{3}_{1}+m^{3}_{2})$. In this reduction formalism
the spherically symmetric
potential $V(r)$ which represents the interaction between the two
particles remains unspecified.

With the shifted angular momentum $l=\bar{l}+\beta$, Eq.(1) becomes
\begin{equation}
-\frac{1}{2\mu}\frac{d^{2}R_{n\ell}(r) }{dr^{2}}+\bar{l}^{2}\left[
\frac{1+(2\beta+1)/\bar{l}+\beta(\beta +1)/\bar{l}^{2}}
{2\mu r^{2}}+\frac{\gamma(r)}{Q}+\frac{E_{n\ell}V(r)}{Q\eta}\right]
R_{n\ell}(r)=(\frac{E^{2}_{n\ell}}{2\eta}+E_{n\ell})R_{n\ell}(r),
\end{equation}
where $\em {n}$ in this paper is the radial quantum number, and Q is a
constant that scales the potential $V(r)$ at large-$l$ limit, and is set for
any specific choice of $l$ and $\em {n}$, equal to $\bar {l}^{2}$ at the end
of the calculations.

Following our earlier work [8], we present here the relevant formulas
obtained in the SLET framework for semi-relativistic motion of a particle
bound in radially symmetric potential. The calculations, and the results are
very simple. The leading-order binding energy is given as:
\begin{equation}
  E_{o}=V(r_{0})-\eta+
  \sqrt{\eta^{2}+\frac{\eta Q}{\mu r_{0}^{2}}}.
\end{equation}
Expanding all the quantities in powers of $1/\bar {l}$ as described in Ref.
[8], one finally gets the second-and third-order corrections $E_{2}$, and
$E_{3}$ of the energy $E_{n\ell}$ as:
\begin{equation}
E_{2}=\frac{Q\alpha_{(1)}}{r_{0}^{2}\left(1+ \frac{E_{0}-V(r_{0})}
{\eta}\right)},
\end{equation}
\begin{equation}
E_{3}=\frac{Q\alpha_{(2)}}{r_{0}^{2}\left(1+\frac{E_{0}-V(r_{0})}{\eta}\right)
},
\end{equation}
where $\alpha_{(1)}$  and $\alpha_{(2)}$ appearing as a correction to the
leading order of the energy expression are given in the appendix of Ref.[11].

 Collecting these terms, and carrying out the mathematics immediately gives
an expression for the energy eigenvalues, that is
\begin{equation}
 E_{n\ell}=E_{0}\;+\;\frac{\alpha_{(1)}}{r_{0}^{2}\left(1\;+\;
 \frac{E_{0}\;-\;V(r_{0})}{\eta}\right)}\;
\;+\;\frac{\alpha_{(2)}}{r_{0}^{2}\left(1\;+\;\frac{E_{0}\;-\;
V(r_{0})}{\eta}\right)\;\bar{l}}\;
\;+\;O\left[\frac{1}{\bar{l}^{2} }\right]\;,
\end{equation}

where $\bar{l}=l-\beta$, $\beta$ is chosen so that the next contribution to
the leading term in the energy eigenvalue series to vanish,
i.e., $E_{1}=0$, which implies that (see Ref.8)
\begin{equation}
\beta=-1/2-\mu(n+1/2)\omega,
\end{equation}
with
\begin{equation}
\omega=\frac{1}{\mu}{\left[3+r_{0}V''(r_{0})/ V'(r_{0})-
\mu r_{0}^{4}V'(r_{0})^{2}/(Q\eta) \right]}^{1/2}\;,
\end{equation}
and $Q$ satisfies
\begin{equation}
 Q=\frac{\mu}{2\eta}\;{\left[r_{0}^{2}V'(r_{0})\right]}^{2}(1+\xi)\;,
\end{equation}\\
with
\begin{equation}
  \xi=\sqrt{1+{[2\eta/r_{0} V'(r_{0})]}^{2}}.
\end{equation}
On the other hand, $r_{0}$ is chosen to minimize $E_{0}$, such that,
\begin{equation}
\frac{dE_{0}}{dr_{0}}\;=\;0
{~~~~\mbox{and}~~~~}~\;\;\;\frac{d^{2}E_{0}}{dr_{0}^{2}}\;>\;0\;,
\end{equation}
and then one can get
\begin{equation}
 1\;+\;2\;\ell\;+\;\mu\;(2\;n\;+\;1)\;\omega\;=\;r_{0}^{2}\;
 V'(r_{0})\;\biggl(\frac{2\mu}{\eta}\;+\;\frac{2\;\xi\;\mu}{\eta}\biggr)^
{1/2},
\end{equation}
which is an explicit equation in $r_{0}$.
It is convenient to summarize the above procedure in the following
steps: (a) Calculate $r_{o}$ from Eq.(12) and substitute it in Eq.(9) to
find $Q$ (b) Substitute $Q$ in Eq.(3) to obtain $E_{o}$. (c) Finally,
one can obtain $E_{o}$ and then calculate $E_{nl}$ from Eq.(6). However, one
is not always able to calculate $r_{o}$ in terms of the potential coupling
constants since the analytical expressions become algebraically complicated,
although straightforward. Therefore, one has to appeal to numerical
computations to find $r_{o}$, and hence $E_{o}$.
\section{ Application, Results and Discussion}
The dilaton $\phi$ is a scalar field predicted  by
superstring theory [12]. The mechanism and the form of the dilaton potential
are unknown, although it is believed that they could be related to
nonperturbative sector
of the theory. Recently it was observed in [5] (see also [6,7]) that a string
inspired coupling of a dilaton  $\phi$ to 4d  $SU(N_c)$ gauge fields
$A_{\mu}=T^{a}A_{\mu}^{a}$, with $T^a$ the $(N_c^2-1)$ $SU(N_c)$ generators,
yields a phenomenologically interesting potential V(r) for the quark-antiquark
interactions. Dick interquark potential was obtained as follow: First
start from the following effective field theory with the dilaton-gluon coupling
$G(\phi)$ and the dilaton potential $W(\phi)$:
\begin{equation}
L(\phi,A)=-{1\over 4G(\phi)} F_{\mu \nu}^{a}F_{a}^{\mu \nu}-{1\over2}
(\partial_{\mu}{\phi})^2+W(\phi)+J_{\mu}^{a}A_{a}^{\mu}
\end{equation}
then construct $G(\phi)$ under the requirement that the Coulomb
problem still possesses analytical solutions. The coupling
$G(\phi)$ and the potential $W(\phi)$ that emerged are:
\begin{equation}
G(\phi)=const.+{f^2\over \phi^2},\qquad
W(\phi)={1\over 2}m^2\phi^2 \qquad
\label{eq:quantum}
\end{equation}
where $f$ is a scale parameter characterizing the strength of the
scalar-gluon coupling and $m$ represents the dilaton mass.

Next, consider the equations of motion of the  fields $A_\mu$ and
$\phi$  and solve them for static point like color source
described by the current density $J_a^{\mu}=\rho_a\eta^{\mu 0}$.
After some straightforward algebra, Dick shows that the
interquark potential $V_{D}(r)$ is given by (up to a color factor),
\begin{equation}
V_D(r)=-{\alpha_s \over r}+gf\sqrt{N_c\over{2(N_c-1)}}\ln[exp(2mr)-1]
\end{equation}
Eq.(15) is remarkable since at large values of $r$ it leads to a
linear confining potential $V_D(r)\sim2gfm\sqrt{N_c\over{2(N_c-1)}}r$.
This derivation provides a challenge to monopole condensations as a
new  quark confinement scenario. Therefore, it is well justified to
dedicate more efforts to the investigation of this effective coupling function
$G(\phi)$ and to the phenomenological application of Dick potential
$V_{D}(r)$.
The method of obtaining results from the theory requires us to choose several
numerical inputs, examination of the previous section shows that there are
five parameters to be calculated, namely, $m_{c}$, $m_{b}$, $m$, $f$ and
$\alpha_s$. A few comments about the choice of the parameters are in order.
The numerical values of the charmed-quark mass $m_{c}=1.89~ GeV$ and the
bottom-quark mass $m_{b}=5.19~GeV$ are chosen so that, the QCD coupling
constant $\alpha_s(\mu)$ at any renormalization scale can be calculated from
the world average experimental value $\alpha_s(m_z)=0.117$ via [13]
\begin{equation}
\alpha_s(\mu)=\frac{\alpha_s(m_z)}
{1-(11-\frac{2}{3}n_f)[\alpha_s(m_z)/2\pi]ln(m_z/\mu)},
\end{equation}
then  one has,
\begin{equation}
\alpha_s(m_c)=0.31, \qquad
\alpha_s(m_b)=0.2. \qquad
\label{eq:quantum}
\end{equation}
For the $b\bar{c}$ quarkonia, we obtain $\alpha_s(4\mu_{bc})=0.22$, where
$\mu_{bc}$ is the reduced mass [14]:
\begin{equation}
\mu_{bc}=\frac{m_bm_c}{m_b+m_c},
\end{equation}
On the other hand, the potential parameters  $m$ and $f$ are considered free
in our analysis and are obtained by fitting the spin-averaged $c\bar{c}$,
and $b\bar{b}$ mesons. An excellent fit with the available experimental
data can be seen to emerge when the following values are assigned
\begin{equation}
m=56.9~ MeV, \qquad
gf\sqrt{\frac{N_c}{2(N_c-1)}}=430~ MeV. \qquad
\label{eq:quantum}
\end{equation}
The results of our calculation for the spin-averaged energy levels of
interest are given in Tables (1,2). In all cases, where comparison with
experiment is possible, agreement is very good. We also present, in Table 3,
the results for the $b\bar{c}$ quarkonia. Our estimate for the $B_c$ mass,
the lowest pseudoscalar S-state of the spectra, is compatible with the
experimental value reported in [15]. As to the higher states masses, they
compare favorably with other predictions based on QCD sum-rules [16] or
other potential models [17]. In conclusion, to the best of our knowledge,
this is the first time that Dick interquark potential of Eq.(15), is tested
and used successfully to fit the spin-averaged  $c\bar{c}$, $b\bar{b}$, and
 $b\bar{c}$ systems. This potential will certainly open a new window in the
potential model applications, essentially those with a Coulomb plus a
confining potential. On the other hand, since a unique theory/scenario for
the dilaton mass is still lacking [18], our estimate for the mass of the
dilaton $m=56.9~ MeV$, resulting from a fit to existing experimental data of heavy quarkonium, lies in the range given in [19]. As suggested by the  authors of Ref.[20], the possibility to identify the dilaton particle to a fundamental scalar, invisible to present day experiment, should not be ruled out.
\clearpage
\vskip 1.5cm
\begin{table}[h]
\caption{Calculated spin-averaged data (in units of GeV) $M_{n\ell}$
of charmonium $c\bar{c}$ energy levels in the semi-relativistic wave
equation using Dick potential}
\vskip 1cm
\begin{center}
\begin{tabular}{|c|c|c||c|c|c|} \hline
 State,$n\ell$ &$M_{n\ell}$, SLET &$M_{n\ell}$, Exp.&
State,$n\ell$ &$M_{n\ell}$, SLET &$M_{n\ell}$, Exp. \\ \hline
1S &3.073 &3.068  &1P&3.548&3.525\\
2S & 3.662 &3.663 &2P&3.871&-  \\
3S&4.027 &4.028 &1D&3.787&3.788   \\
\hline
\end{tabular}
\end{center}
\end{table}
\vskip 1cm
\begin{table}[h]
\caption{Calculated spin-averaged data (in units of GeV) $M_{n\ell}$
of bottomonium $b\bar{b}$ energy levels in the semi-relativistic wave
equation using Dick potential}
\vskip 1cm
\begin{center}
\begin{tabular}{|c|c|c||c|c|c|} \hline
 State,$n\ell$ &$M_{n\ell}$, SLET &$M_{n\ell}$, Exp.&
State,$n\ell$ &$M_{n\ell}$, SLET &$M_{n\ell}$, Exp. \\ \hline
1S &9.450 &9.446  &1P&9.903&9.900\\
2S & 10.014 &10.013 &2P&10.206&10.260  \\
3S&10.292 &10.348 &1D&10.129&-   \\
\hline
\end{tabular}
\end{center}
\end{table}
 \vskip 1cm
\begin{table}[h]
\caption{Calculated spin-averaged data (in units of GeV) $M_{n\ell}$
of $b\bar{c}$ energy levels in the semi-relativistic wave
equation using Dick potential}
\vskip 1cm
\begin{center}
\begin{tabular}{|c|c|c||c|c|c|} \hline
 State,$n\ell$ &$M_{n\ell}$, SLET &$M_{n\ell}$, Exp.&
State,$n\ell$ &$M_{n\ell}$, SLET &$M_{n\ell}$, Exp. \\ \hline
1S &6.322 &6.40 $\pm0.39$$\pm0.19$&1P&6.767&-\\
2S & 6.876 &- &2P&7.072&-  \\
3S&7.161 &- &1D&6.994&-   \\
\hline
\end{tabular}
\end{center}
\end{table}


\clearpage
\begin{thebibliography}{99}
\bibitem{R1} W. Lucha and F. F. Sch\"{o}berl, and D. Gromes Phys. Rep.
{\bf C200}  (1991) 127.
\bibitem{R2} K. G. Wilson, Phys. Rev. {\bf D10}  (1974) 2445; Phys. Rep.
{\bf C23} (1976) 331; J. Kogut and L. Susskind Phys. Rev. {\bf 
D11}(1975) 395.
\bibitem{R3} K. Johnson and C. Thorn, Phys. Rev. {\bf D13} (1976)
1934.
\bibitem{R4} E. Eichten, K. Gottfrid T. Kinoshita, K. D. Lane and T. M. Yan,
Phys. Rev. {\bf D17} (1978) 3090; ibid. 21,  (1980) 313 (E); ibid.
21, (1980) 203.
\bibitem {R5} R. Dick, Eur. Phys. Journal. {\bf C6 }(1999) 701; Phys. lett.
{\bf B397} (1997) 193; Phys. lett. {\bf  B409} (1997) 321.
\bibitem{R6} M. Chabab, R. Markazi, and E. H. Saidi, Eur. Phys. Journal
{\bf C13}, (2000) 543; M. Chabab, "Comments on confinement from
the dilaton-gluon coupling in QCD", to appear in the Proceedings
of the International Conference on Quark Confinement and the
Hadron spectrum, (Vienna, 03-08 July 2000), eds. W. Lucha and F.
Sch\"{o}berl (World Scientific, River Edge, N. J. 2001);
hep-th/0009115, LPHEA/00-02.
\bibitem {R7} R. Dick and L.P. Fulcher,  Eur. Phys. Journal {\bf C397}
(1999) 271.
\bibitem {R8} T. Barakat, Int. J. Mod. Phys. {\bf A}, (in print);
math-ph/0004026; T. Barakat, "Application of the Shifted-l Expansion method
to B, and D Meson Leptonic Decay Constants in the Semi-Relativistic Wave
Equation", to appear in the Proceedings of the International Conference on
Quark Confinement and the Hadron Spectrum, (Vienna, 03-08 July 2000), eds. W.
Lucha and F. Sch\"{o}berl (World Scientific, River Edge, N. J. 2001).
\bibitem {R9} D. Sung Hwang, and G. Hee Kim, Phys. Rev. {\bf D53}
 (1995) 3659; D. Sung Hwang et al., Phys. Rev. {\bf D53} (1996) 4951.
\bibitem {R10} F. Brau, J. Math. Phys. {\bf 39} (1998) 2254.
\bibitem {R11} T. Barakat, M. Odeh  and O. Mustafa, J. Phys. {\bf A31},
 (1998) 3469.
\bibitem {R12} M. Green, J. Schwartz, and E. Witten, « Superstring Theory,
Cambrige University Press (1987).
\bibitem{R13} Keiji Igi, and Seiji Ono, Phys. Rev. {\bf D32} (1985) 232.
\bibitem{R14}  L. Motyka, and K. Zalewski, Eur. Phys. Journal. {\bf C4}
(1998) 107.
\bibitem{R15} F. Albe et al., CDF Collaboration,  Phys. Rev. Lett. {\bf 81}
 (1998) 2432.
\bibitem{R16} M. Chabab, Phys. Lett.{\bf B325}  (1994) 205; Talk presented in
the 7th International Conference on Hadon Spectroscopy, Upton, NY, 25-30
August, 1997; AIP Proceeding no {\bf 432}, pp. 856-860.
\bibitem{R17} E. Eichten and C. Quigg, Phys. Rev. {\bf D49} (1994) 584
(and references therein);  S.S. Gershtein, V.V. Kiselev, A.K. Likhoded and
A.V. Tkabladze, Phys. Usp. {\bf 38} (1995) 1; N. Brambilla and A. Vairo,
{\bf hep-ph/0002075}.
\bibitem{R18} A.B. Kaplan and M.B. Wise,
''Coupling of a light dilaton and violations of the equivalence principle'',
{\bf hep-th/0008116}.
\bibitem{R19} M. Gasperini, Phys. Lett. {\bf  B327} (1994) 214; Y.M. Cho and Y.Y.
Keum, Mod. Phys. Lett. {\bf  A3} (1998) 108.
\bibitem{R20}M. Bando, K.I. Matumoto, K. Yamawaki, Phys. Lett. {\bf B178} (1986)
308; E. Halyo, Phys. Lett. {\bf B271} (1991) 415.
\end{thebibliography}
\end{document}